\documentclass[journal,twoside]{IEEEtran}
\usepackage{times}
\usepackage{soul}

\usepackage[hidelinks]{hyperref}
\usepackage[utf8]{inputenc}
\usepackage[small]{caption}
\usepackage{graphicx}
\usepackage{amsmath}
\usepackage{booktabs}
\usepackage{algorithm}
\usepackage{algorithmic}
\usepackage{multirow}
\usepackage{mathrsfs}
\usepackage{amsthm}
\usepackage{amsfonts}
\usepackage{stfloats}
\usepackage{color}
\usepackage{flushend}
\newtheorem{myDef}{Definition}

\begin{document}

\title{Modeling and Understanding Ethereum Transaction Records via A Complex Network Approach}
 
\author{Dan~Lin, Jiajing~Wu,~\IEEEmembership{Senior Member,~IEEE,}
	Qi~Yuan, and~Zibin~Zheng,~\IEEEmembership{Senior Member,~IEEE}
	\thanks{Manuscript received December 26, 2019; accepted January 16, 2020. Date of publication January 21, 2020; date of current version November 4, 2020. This work was supported in part by the National Key Research and Development Program under Grant 2016YFB1000101, in part by the National Natural Science Foundation of China under Grant 61973325 and Grant 61503420, and in part by the Guangdong Province Universities and Colleges Pearl River Scholar Funded Scheme (2016). \emph{(Corresponding Author: Jiajing Wu.)}
	}
	\thanks{The authors are with the School of Data and Computer Science, and the National Engineering Research Center of Digital Life, Sun Yat-sen University, Guangzhou 510275, China. (Email: wujiajing@mail.sysu.edu.cn)}% <-this % stops a space
	\thanks{Digital Object Identifier 10.1109/TCSII.2020.2968376}
	}

\markboth{IEEE Transactions on Circuits and Systems II: Express Briefs,~vol.~67, No.~11, November~2020}%
{Lin \MakeLowercase{\textit{et al.}}: Modeling and Understanding Ethereum Transaction Records}

\maketitle

\begin{abstract}
As the largest public blockchain-based platform supporting smart contracts, Ethereum has accumulated a large number of user transaction
records since its debut in 2014. Analysis of Ethereum transaction records, however, is still relatively unexplored till now. Modeling the transaction records as a static simple graph, existing methods are unable to accurately characterize the temporal and multiplex features of the edges. In this brief, we first model the Ethereum transaction records as a complex network by incorporating time and amount features of the transactions, and then design several flexible temporal walk strategies for random-walk based graph representation of this large-scale network. Experiments of temporal link prediction on real Ethereum data demonstrate that temporal information and multiplicity characteristic of edges are indispensable for accurate modeling and understanding of Ethereum transaction networks.
\end{abstract}


\IEEEpeerreviewmaketitle

\section{Introduction}
\IEEEPARstart{B}{eing} the largest public blockchain-based platform supporting smart contracts, Ethereum~\cite{wood2014ethereum} has attracted wide attention recently. To facilitate the implementation of smart contracts, Ethereum introduces the concept of \textit{account}, which is formally an address. The corresponding cryptocurrency on Ethereum, known as \textit{Ether}, can be transferred between accounts and used to compensate participant mining nodes. Current research about Ethereum mainly focus on security and performance issues of the blockchain technology~\cite{8530775}, \cite{chen2019cooperative}, \cite{chen2019secure}, however, the interative relationship between users and smart contracts is relatively unexplored till now. 

Complex network is a universal tool to analyze real work systems from various fields~\cite{7931718}, \cite{chen2019Comparative}, \cite{zhou2019Importance}, and it has been employed to model and analyze the huge transaction networks of blockchain-based systems.
In 2018, Chen \textit{et al.}~\cite{chen2018understanding} conducted the first systematic study to characterize Ethereum and obtain new observations with various metrics. Alqassem \textit{et al.}~\cite{8575170} found the anti-social properties of Bitcoin system via transaction network data analysis.
In existing studies, transaction records are modeled as a simple graph, where multiple transactions between a pair of addresses are merged as a one-time transaction in the graph construction procedure.

Different from other large-scale complex networks, each edge in the Ethereum transaction network represents a particular Ether transaction, and thus contains some unique information such as the direction, amount value and timestamp of a particular transaction. It is essential to incorporate the aforementioned information for accurate modeling, characterization, and understanding of transaction network data. In addition, multiple transactions between two users are expected and it is more comprehensive to model a transaction network as a \textit{multidigraph} rather than a simple graph. In graph theory, a multigraph (in contrast to a simple graph) is a graph which is permitted to have self-loops and multiple edges (also called parallel edges). A multidigraph is a directed multigraph. At present, there is little research on data mining of multidigraphs, and existing methods modeling the transaction records as simple graph are inadequate to characterize the temporal and multiple features of transaction networks.

Random walk mechanism has been widely demonstrated to be an effective technique to measure local similarity of networks for a variety of  domains and tasks~\cite{spitzer2013principles,Liu_2010,Rosvall1118}.
Besides, random walk has also become a popular sampling method for the problem of graph representation, which aims to represent node features of large networks in a low dimensional space for graph analysis and mining~\cite{cai2018comprehensive}. Most existing methods of graph representation are designed for mining social networks and few studies have considered the scenarios of transaction data mining as most financial transaction data is not public.
However, traditional methods for graph representation in social networks cannot be directly applied to analyze financial transaction networks, as the existing methods~\cite{cai2018comprehensive} ignore the multiplicity and temporal attributes of edges in transaction networks.

In this brief, we conduct the \textit{first} work to understand Ethereum transaction records via random-walk based graph representation learning. In particular, we first model the Ethereum transaction records as a \textit{Temporal Weighted Multidigraph}, and propose novel graph sampling and representation methods considering the temporal and weighted information of the graph. After that, we conduct link prediction experiments on realistic Ethereum data to evaluate the effectiveness of the proposed graph modeling and mining methods.

\section{Framework}

In order to analyze the Ethereum transaction records comprehensively, we propose a general framework which includes four main procedures: 
(a) Data collection. We fetch transaction records of objective accounts from Ethereum through API of https://etherscan.io/ (A block explorer and analytics platform for Ethereum); (b) Network construction. Based on collected transaction data, we construct a \textit{Temporal Weighted Multidigraph} to represent the relationship of Ether transfer between accounts; (c) Random-walk based graph representation. We employ the graph representation algorithm based on designed random walk strategies to learn richer node representations; (d) Link prediction for evaluation. We conduct experiments on temporal link prediction problem to evaluate the effectiveness of our framework on Ethereum transaction networks.

\section{Data Collection}

Accounts on Ethereum can be divided into two categories, i.e., external owned accounts (EOA) which are similar to general bank accounts~\cite{chenweili2018blockchaindata}, and smart contract accounts which are source code files. In this work, we focus on the transactions among EOAs for the reason that the Ether transfer records between them are publicly available on the blockchain. Besides, we only include the successful transactions with non-zero amount value into our dataset.
	
\begin{figure}[htbp]	
	\centering	
	\vskip -0.2in
	\includegraphics[scale=0.3]{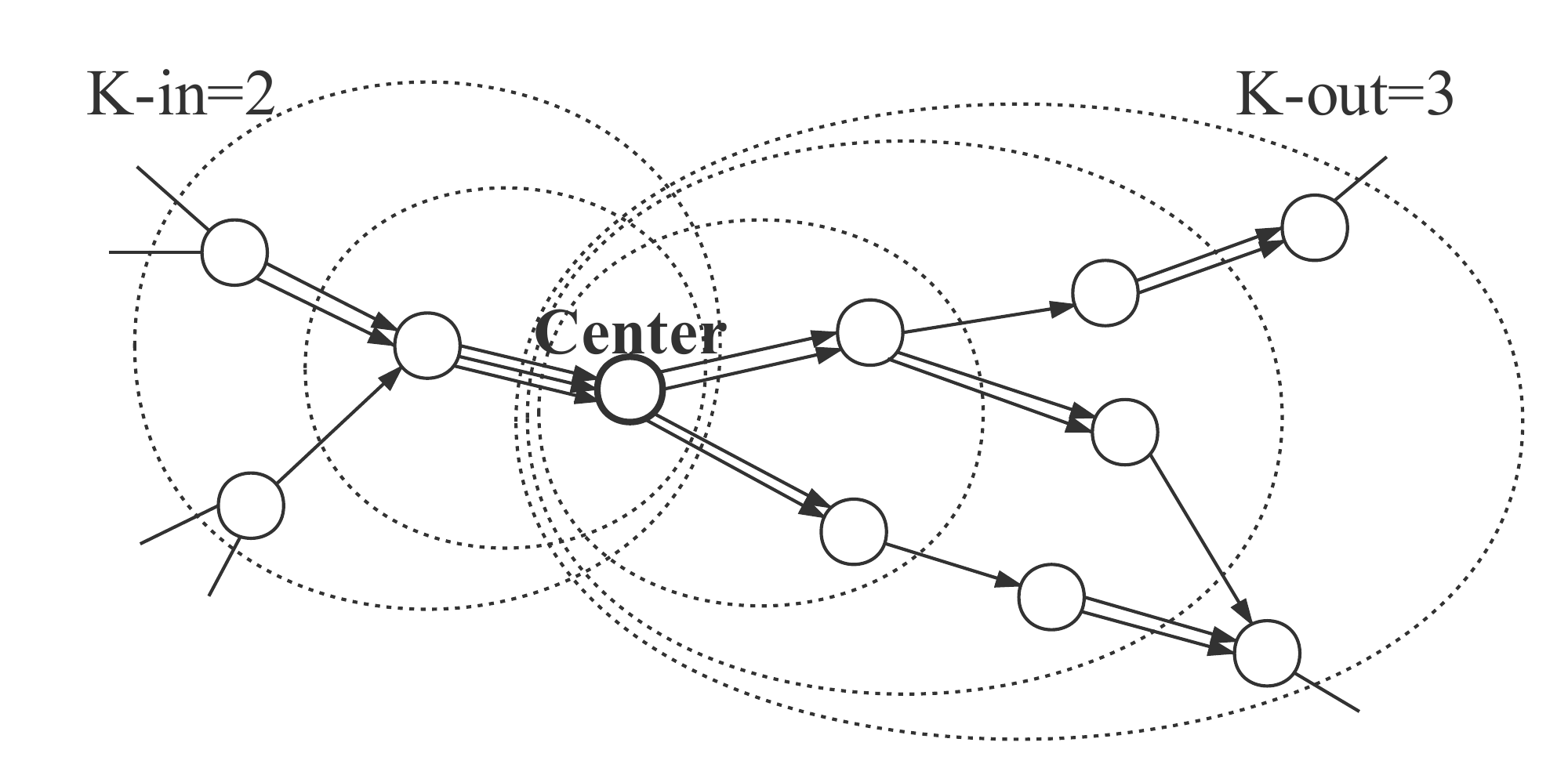}	
	\caption{A schematic illustration of a directed $K$-order subgraph.}	
	\label{fig:korder}	
	\vskip -0.1in
\end{figure}

Since it is extremely time-consuming to process the whole Ethereum transaction network with more than two million EOAs~\cite{chen2018understanding}, here we select a number of objective accounts and then obtain their transaction data through APIs of Etherscan. As shown in Fig.~\ref{fig:korder}, centered by a particular objective account, we obtain a directed $K$-order subgraph. $K$-in and $K$-out are two parameters to control the depth of sampling inward and outward from the center, respectively.
	
On Ethereum, various related information of Ether transactions is stored as data packages. In details, the \textit{TxHash} field is a unique identification of a transaction, the \textit{Value} field in a transaction refers to the amount of money transferred, and the~\textit{Timestamp} field indicates when the transaction happens. Besides, the \textit{From} and \textit{To} fields denote the sender and recipient of the transaction. With the collected four-tuples $(From, To, Value, Timestamp)$, we can construct a temporal weighted multidigraph.
	
In this brief, we conduct the first work to understand Ethereum transaction records from a complex network perspective, and we will open our collected dataset from Ethereum for future studies.

\section{Network Construction}
\label{NetworkConstruction}

\begin{figure}[tbp] 
\centering
\includegraphics [scale=0.85]{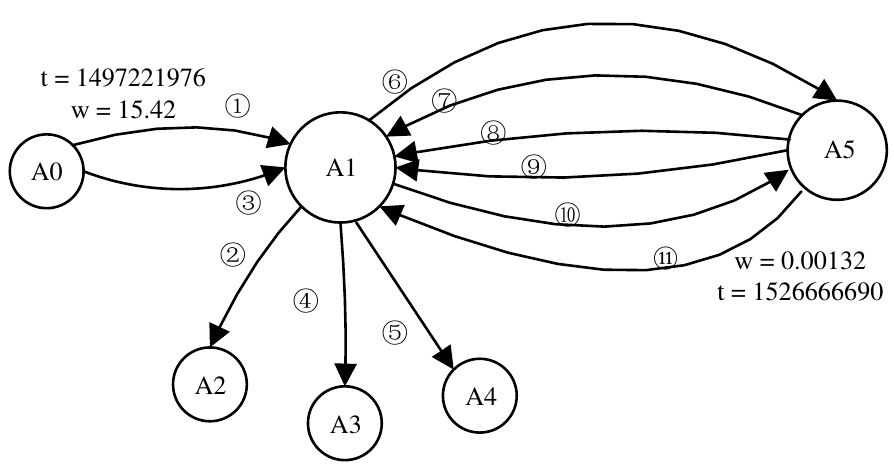}
\vskip -0.05in
\caption{An illustration for the Ethereum transaction network. Nodes are labeled by account addresses. Edges are attached by timestamp $t$ and amount value $w$, and indexed in the increasing order of $t$. }
\label{fig:micro_transaction_network}
\vskip -0.15in
\end{figure}

Before network construction, we first discuss some realistic rules and features of transaction networks like the Ethereum: 
(1) Transaction networks evolve continuously over time with additions of links, which is overlooked in most of the existing random-walk based graph algorithms;
(2) The practical meaning of connections between accounts is not a one-off established relationship but a time-dependent event. Hence multiple edges need to be considered in transaction network modeling; 
(3) Unlike social networks, random walks on Ethereum transaction network are concrete, which represent realistic money transfer flows;
(4) The amount value of a transaction reflects the similarity between the two involved accounts to some extent. In most cases, the larger amount of transaction, the closer relationship between two accounts.
Fig.~\ref{fig:micro_transaction_network} demonstrates a microcosm of transaction activities on Ethereum.

Ether transfer is one of the major activities happening on Ethereum. Here we abstract an Ether transfer transaction as a four-tuple (\textit{src}, \textit{dst}, \textit{w}, \textit{t}), which means the sender \textit{src} transfers \textit{w} Ether to the recipient \textit{dst} at time \textit{t}. To investigate the Ether transfer on Ethereum, we abstract the Ethereum transaction network as a \textit{Temporal Weighted Multidigraph}:
	
\begin{myDef}[Temporal Weighted Multidigraph (TWMDG)]
	Given a graph $G=(V,E)$, let $V$ be the set of nodes and $E$ be the set of edges. Each edge is unique and is represented as $e = (u, v, w, t)$, where $u$ is the source node, $v$ is the target node, $w$ is the weight value and $t$ is the timestamp. For the sake of simplicity, we define mapping functions $Src(e)=u$, $Dst(e)=v$, $W(e)=w$, $T(e)=t$ for $\forall e\in E$. 
\end{myDef}
	
Based on collected four-tuples from Ethereum transaction records, we can build a TWMDG where each node represents a unique account and each edge represents a unique Ether transfer transaction.

\section{Random-Walk Based Graph Representation}

\begin{figure}
	\centering	
	\includegraphics[scale=0.75]{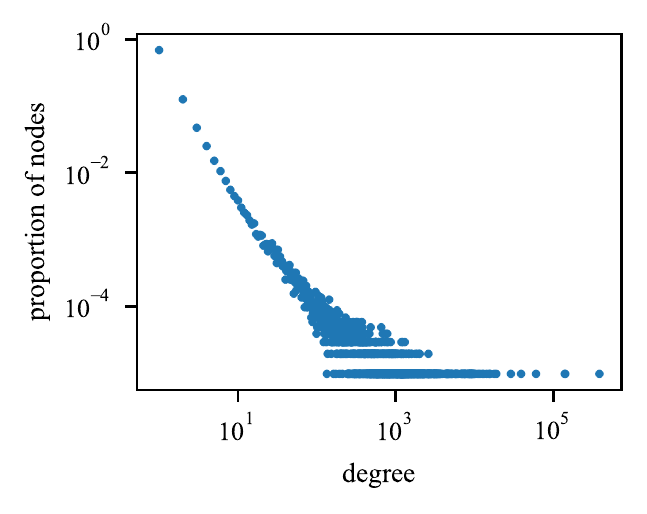}
	\vskip -0.12in
	\caption{The distribution of nodes appearing in Ethereum transaction subgraph follows a power-law.}
	\label{fig:degree}
	\vskip -0.2in
\end{figure}

Random-walk based graph representation methods have been proved to be scalable and effective for large graphs. As realistic transaction networks usually have a huge network size, we consider the random walk based method for graph analysis in this area.
Before utilizing graph representation algorithm, we investigate the degree distribution of Ethereum transaction network, and find that the distribution of nodes in both the entire graph and the subgraph of Ethereum follow power-law (Fig.~\ref{fig:degree} plots the distribution of a subgraph.). As discussed in a pioneering study~\cite{perozzi2014deepwalk}, random-walk based sampling methods can well preserve structural properties of a graph with a power-law degree distribution. Inspired by this study, we employ graph representation method based on random walk to extract information from the Ethereum transaction network.
	
However, for the transaction networks discussed here, existing methods that ignore temporal information may sample a large number of invalid transaction sequences to derive node representations.	
For example, in Fig.~\ref{fig:micro_transaction_network}, $\{A5, A1, A2\}$ is a possible random walk sequence in traditional methods DeepWalk~\cite{perozzi2014deepwalk} and node2vec~\cite{grover2016node2vec}, but it is not practical in a temporal graph as the transaction from $A1$ to $A2$ happens earlier.
%The transaction $\Large{\textcircled{\small{2}}}$  from $A1$ to $A2$ cannot contain any fortune and information interacted in the transaction from $A5$ to $A1$. 
While in a recent work reported in~\cite{Nguyen:2018:CDN:3184558.3191526}, although temporal information is considered, the existence of multiple edges between points is neglected. For instance, the temporal walk from $A0$ to $A1$ is represented as a sequence of nodes $\{A0, A1\}$.
However, whether $A2$ is possible for the next walk depends on whether the transaction path $\textcircled{\small{1}}$ or $\textcircled{\small{3}}$ is sampled by the previous walk from $A0$ to $A1$. 
	
In this work, we represent an $l$-length temporal walk as a sequence of $l$ nodes together with a sequence of $(l-1)$ edges traversed in non-decreasing timestamps. This kind of temporal walk represents an actually feasible path for money flow in the transaction network. Therefore, the proposed method is expected to learn more meaningful and accurate time-dependent node representations that can capture more comprehensive properties from dynamic transaction networks.
For a temporal weighted multidigraph discussed here, we define the concept of a \textit{Temporal Walk} as follows:
	
\begin{myDef}[Temporal Walk]
	In TWMDG, a temporal walk from node $v_1$ to $v_l$ is an $l$-length path traversed in non-decreasing timestamps. Such a temporal walk is represented as a sequence of $l$ nodes $ walk_n=\{v_1, v_2, ..., v_l\}$ together with a sequence of $(l-1)$ edges $ walk_e=\{e_1, e_2, ..., e_{l-1}\}$, where $Src(e_i)=v_i$, $Dst(e_i)=v_{i+1}$ $(1\leq i \leq(l-1))$, and $T(e_i)\leq T(e_{i+1})$ $(1\leq i \leq (l-2))$. We define that nodes $u$ and $v$ are temporally connected if there exists a temporal path from $u$ to $v$.
\end{myDef}

\begin{figure}[h] 
	\centering
	\vskip -0.15in
	\includegraphics [scale=0.4]{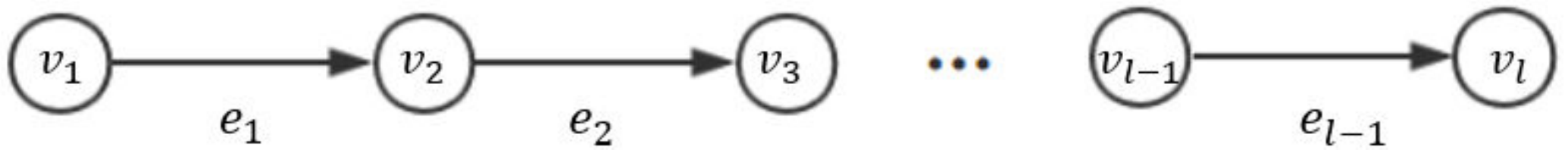}
	\caption{Illustration of an $l$-length temporal walk }
	\label{fig:l-walk}
	\vskip -0.15in
\end{figure}
	
In order to sample valid random walks obeying the temporal constraint, we introduce a new concept called \textit{Temporal Successive Edges} in TWMDG.
		
\begin{myDef}[Temporal Successive Edges]
	Given a temporal weighted multidigraph $G=(V,E)$, the temporal successive edges of a node $u$ at time $t$ is defined as follows:
	$$ L_t(u) = \{ e~|~Src(e)=u, T(e)\geq t \} $$
\end{myDef}
For instance, in Fig.~\ref{fig:micro_transaction_network}, let $t=T(e_{\scriptsize{\textcircled{\tiny{5}}}})$, then $L_t(A1)=\{ e_{\scriptsize{\textcircled{\tiny{5}}}}, e_{\scriptsize{\textcircled{\tiny{6}}}}, e_{\scriptsize{\textcircled{\tiny{10}}}}\}$. The set of temporal successive edges plays the role of candidates for walkers to select possible successors.

Apart from the temporal constraint, we further develop biased searching strategies by considering more detailed transaction information. 
For the Ethereum transaction network discussed here, we abstract the transaction time and amount as the temporal and weighted information of a TWMDG. 
Consider a random walk that just traversed edge $e_{i-1}$, and is now stopping at node $v_i$ at time $t=T(e_{i-1})$. The next node $v_{i+1}$ of the random walk is decided by selecting a temporally valid edge $e_i$. We describe different sampling biases by formulating the selection probability for each temporal successive edge $e\in L_t(v_i)$.

From the perspective of temporal domain, we consider both unbiased and biased sampling strategies as follows.
\begin{itemize}
	\item \textbf{Temporal Unbiased Sampling (TUS)}. This is the default setting in the time domain, which assumes that each temporal successive edge $e\in L_t(v_i)$ of node $v_i$ at time $t$ has the same probability to be selected:
	\begin{equation}
		\label{equ:time_uniform}
		P_T(e)=\frac{1}{|L_t(v_i)|}.
	\end{equation}
		
	\item \textbf{Temporal Biased Sampling (TBS)}. 
	For financial transaction networks, the similarity between accounts is time-dependent and dynamic. 
		
	On the one hand, a pair of accounts with frequent interactions are supposed to have a stronger relationship. Therefore, we let $\eta_-: \mathbb{R}\rightarrow \mathbb{Z}^+$ be a function that maps the timestamps of edges to a descending ranking. In this case, each edge $e\in L_t(v_i)$ will be assigned with a selection probability:
	\begin{equation}
		\label{equ:time_close_linear}
		P_T(e)=\frac{\eta_-(T(e))}{\sum_{e'\in L_t(v_i)}~\eta_-(T(e'))}.
	\end{equation}	
	where $T(e)$ denotes the timestamp of the edge $e$.
	This sampling method biases the selection towards edges that are closer in time to the previous edge.
		
	On the other hand, sampling the interactions among accounts in a large time interval may also be important for different domains of networks for the purpose of preserving global similarity in time domain. For such scenarios, we propose another strategy that favors edges appearing later to the previous timestamp. Let $\eta_+:\mathbb{R}\rightarrow \mathbb{Z}^+$ be a function that maps the timestamps of edges to an ascending ranking. The probability of selecting each edge $e\in L_t(v_i)$ can be given as:
	\begin{equation}
		\label{equ:time_far_linear}
		P_T(e)=\frac{\eta_+(T(e))}{\sum_{e'\in L_t(v_i)}~\eta_+(T(e'))}.
	\end{equation} 
		
\end{itemize}

Apart from the transaction time, the amount values of the edges (edge weights) also plays an essential role in financial transaction networks. In the following, we present unbiased and biased strategies from a weighted domain.
\begin{itemize}
		\item \textbf{Weighted Unbiased Sampling (WUS)}. Similar to TUS, this is the default setting in the amount domain and each edge $e\in L_t(v_i)$ has the same probability to be sampled:
		\begin{equation}
		\label{equ:amount_uniform}
		P_W(e)=\frac{1}{|L_t(v_i)|}.
		\end{equation}
		
		\item \textbf{Weighted Biased Sampling (WBS)}. As illustrated in the Introduction, the weight value of each transaction indicates the significance of interactions between the two accounts involved. For most instances, a higher value of transaction amount implies a larger similarity between the two accounts. Thus each edge $e\in L_t(v_i)$ can be assigned the selection probability:
		\begin{equation}
		\label{equ:amount_raw}
		P_{W}(e)=\frac{W(e)}{\sum_{e’\in L_t(v_i)} W(e')}.
		\end{equation}
		
		To prevent the extreme situation where edges with small weights would never be sampled, we consider a linear mapping function to weakens the effects of edge weights. Thus we have
		\begin{equation}
		\label{equ:amount_linear}
		P_W(e)=\frac{\eta_+(W(e))}{\sum_{e'\in L_t(v_i)} \eta_+(W(e'))}.
		\end{equation}
		
\end{itemize}

Furthermore, we combine the aforementioned sampling probabilities from both temporal and weighted domains, i.e., $P_T$ and $P_W$, by $P(e) = P_T(e)^\alpha P_W(e)^{(1-\alpha)}  (0\leq\alpha\leq1)$ for  $\forall e\in L_t(v_i)$. Here $\alpha=0.5$ is the default value for balancing between time domain and amount domain. 
After the random walk generation with the temporal constraint and flexible biased strategies, the second part is an update procedure based on SkipGram~\cite{Mikolov2013Efficient,mikolov2013distributed}, which learns node representations as a maximum likelihood optimization problem.
	
\begin{figure}
	\centering	
	\includegraphics[scale=0.43]{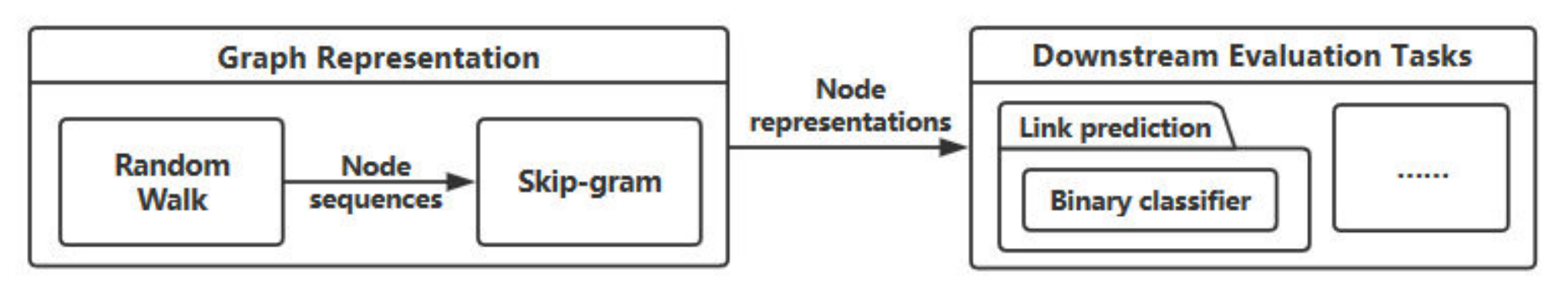}	
	\caption{The illustration of Ethereum transaction network modeling and analysis.}	
	\label{fig:flow}	
	\vskip -0.2in
\end{figure}

\section{Link Prediction for Evaluation}

In the previous section, we discuss different walking strategies for graph representation learning, which aims to preserve network properties for downstream tasks. Usually, link prediction and node classification are two common downstream tasks for evaluation of graph representation learning~\cite{cai2018comprehensive}. Due to the unavailability of ground truth of Ethereum accounts, we evaluate our temporal-random-walk-based graph embedding by link prediction on realistic Ethereum data, as Fig.~\ref{fig:flow} shows.
In fact, link prediction is also a valuable issue in blockchain systems. A series studies on Ethereum have witnessed various illegal behaviors or scams such as phishing, Ponzi~\cite{Chen:2018:DPS:3178876.3186046}, money laundry~\cite{6805780}, etc., and the temporal link prediction can help people track the real-time process of Ether flow or prevent illegal users’ cashing process~\cite{8740574}.

The task of link prediction aims to predict the occurrence of links in a given graph on the basis of observed information. In the temporal link prediction problem, unlike the static link prediction where links have no timestamp, we use the existing links in the past (50\% of the earlier edges) as the training data to predict the occurrences of links in the future (the remainder).
Shown in TABLE~\ref{tab:4dataset}, we randomly select center accounts and collect three subgraphs with different size ($K$-in = $K$-out = $K$) from Ethereum for learning node embedding vectors $\Phi(v)$ for $\forall v \in V$ via random-walk based graph representation learning.
Next, we use a binary classifier for supervised link prediction, where node pairs $(src, dst)$ existing in the training graph act as positive samples of the training set. We randomly sample an equal number of node pairs with no link as negative samples. Then we obtain features of a directed link from nodes $v_i$ to $v_j$ by concatenating their node representations, i.e., $F_{i,j}=[\Phi(v_i), \Phi(v_j)]$. If $i\neq j$, $F_{i,j}\neq F_{j,i}$. Finally, we train a support vector classifier to classify the links in the test set (50\% of the later edges).

\begin{figure*}
	\centering	
	\includegraphics[scale=0.82]{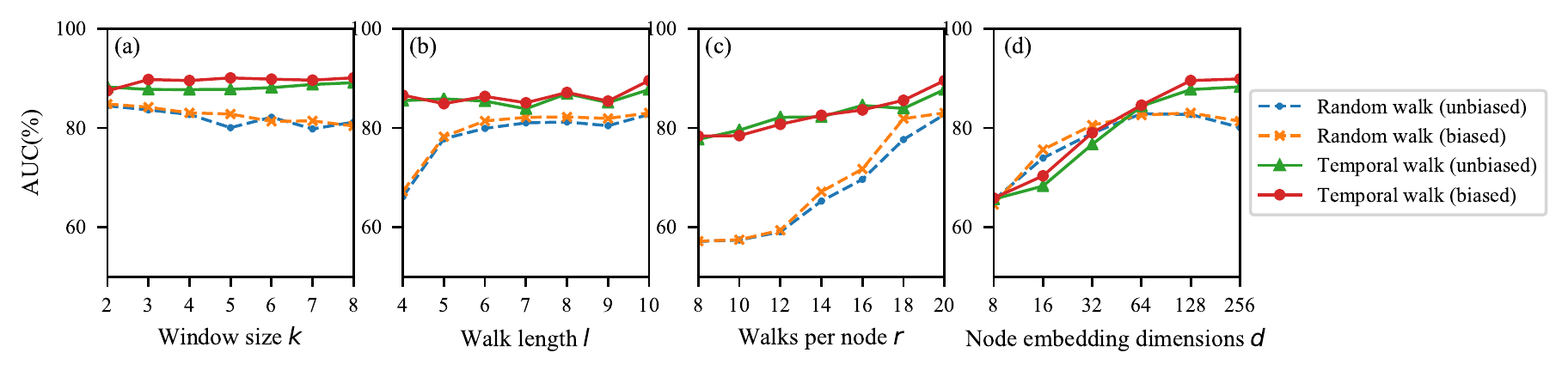}	
	\vskip -0.15in
	\caption{Performance in terms of Area Under Curve (AUC) under varying hyperparameters, when (a) fixing $l=10$, $r=20$, $d=128$, and varying $k$ from 2 to 8; (b) fixing $k=4$, $r=20$, $d=128$, and varying $l$ from 4 to 10; (c) fixing $k=4$, $l=10$, $d=128$, and varying $r$ from 8 to 20; (d) fixing $k=4$, $l=10$, $r=20$, and varying $d$ from 8 to 256. }	
	\label{fig:parameter}	
	\vskip -0.2in
\end{figure*}

\newcommand{\tabincell}[2]{\begin{tabular}{@{}#1@{}}#2\end{tabular}}
\begin{table}[tbp] 
	\centering 
	\scriptsize 
	\begin{tabular}{ccccc}
		\toprule
		
		Dataset & Address of Center Node & $K$ & \#Accounts & \#Transactions \\ \midrule
		
		EthereumG1 & \tabincell{c}{0x51faeda318982f439e8\\0012fb45d2b017ddccdbe} & 3 & 3,832 & 208,927 \\ \midrule
		EthereumG2 & \tabincell{c}{0x5e247060f48eeb6436\\7250ed03ff5091bba47fd1} & 4 & 10,628 & 208,533 \\ \midrule
		EthereumG3 & \tabincell{c}{0x51faeda318982f439e8\\0012fb45d2b017ddccdbe} & 4 & 26,175 & 677,785 \\ 
		\bottomrule
	\end{tabular}
	\vskip -0.05in
	\caption{Statistics of datasets}
	\label{tab:4dataset}
	\vskip -0.1in
\end{table}

\begin{table}[tbp]
	\centering
	\scriptsize 
	\begin{tabular}{p{8em}cccccc}
		\toprule
		\multirow{2}[4]{*}{\textbf{Metrics(\%)}} & \multicolumn{2}{c}{\textbf{EthereumG1}} & \multicolumn{2}{c}{\textbf{EthereumG2}} & \multicolumn{2}{c}{\textbf{EthereumG3}} \\
		\cmidrule{2-7}    \multicolumn{1}{c}{} & \textbf{AUC} & \textbf{AP} & \textbf{AUC} & \textbf{AP} & \textbf{AUC} & \textbf{AP} \\
		\midrule
		\textit{Directed graph + unbiased walk} & 82.71  & 76.69  & 85.91  & 82.13  & 79.92  & 77.72  \\
		\textit{Directed graph + biased walk} & 83.03  & 76.94  & 86.30  & 82.47  & 82.20  & 79.99  \\
		\textit{TWMDG + unbiased temporal walk} & 87.73  & 83.73  & 92.85  & 90.29  & 93.00  & 90.78  \\		
		\textit{TWMDG + biased temporal walk} & \textbf{89.55} & \textbf{85.58} & \textbf{93.36} & \textbf{90.94} & \textbf{93.83}  & \textbf{91.89} \\
		\bottomrule
	\end{tabular}
	\vskip -0.05in
	\caption{ Performances of different graph modeling and walking methods for link prediction.}
	\label{tab:performances}
	\vskip -0.2in
\end{table}

\paragraph{Settings} 

In the experiments, we compare our proposed walking strategies with directed DeepWalk~\cite{perozzi2014deepwalk} and directed node2vec~\cite{grover2016node2vec} implemented by OpenNE~\cite{openne}.
For the random walk based representation methods, we have several hyperparameters: the node representation dimension $d$, the size of window $k$, the length of walk $l$, and walks per node $r$. In general, we set $d=128$, and $k=4$. Specifically, we set $r=20$, $l=10$ for EthereumG1, $r=10$, $l=10$ for EthereumG2, $r=10$, $l=20$ for EthereumG3.
For the purpose of comparisons, we consider two widely discussed random walk based methods, namely DeepWalk~\cite{perozzi2014deepwalk} and node2vec~\cite{grover2016node2vec}, labeled as \textit{Directed graph+unbiased walk} setting and \textit{Directed graph+biased walk} settings in Table~\ref{tab:performances}, respectively. Besides, we apply temporal walk with TUS and WUS for \textit{TWMDG+unbiased temporal walk} and temporal walk with TBS and WBS for \textit{TWMDG+biased temporal walk} ($\alpha=0.5$).

\paragraph{Results}	
Table \ref{tab:performances} compares the performance of various methods on temporal directed link prediction in terms of Area Under Curve (AUC) and Average Precision (AP). Modeling the transaction data as a TWMDG without any bias overwhelmingly outperforms traditional directed graph. Temporal walk with biases of both time and amount domains leads to better performance than unbiased temporal walk.

\paragraph{Dicussions of Results}
The results manifest that the temporal information as well as the multiplicity characteristic of edges in TWMDG are very important and meaningful for analysis and understanding of financial transaction networks. The comparisons also demonstrate that the rich information from time and amount domains does help us obtain a more comprehensive representation for predictive tasks.

\paragraph{Parameter Analysis}
	
To further illustrate the superiority of biased temporal walk, we compare the performance on EthereumG1 with varying value of node representation dimension $d$, walk length $l$, walks per node $r$ and window size $k$. Results in Fig.~\ref{fig:parameter} demonstrate that: (1) Temporal walk with or without additional biases consistently outperform unbiased and biased random walk under different settings of $k$, $l$, $r$; (2) Random-walk based methods are more sensitive to two hyperparameters, walk length $l$ and walks per node $r$, while temporal-walk based methods can always achieve promising results with a wide range of both $l$ and $r$; (3) Interestingly, with an increase of $d$, the performance of temporal-walk based methods monotonically improves but performance of random-walk based methods degrades with $d$ larger than 64, which implies that temporal walk can embed richer helpful information and thus requiring a larger value of $d$ for data representation.

\section{Conclusion and Discussion}

In this brief, we proposed to model the Ethereum transaction network as a new defined network model called temporal weighted multidigraph, and refine the definition of temporal walk considering the network dynamics and the multiplicity of edges to obtain graph representation more accurately. Experiments on the task of temporal link prediction demonstrated the effectiveness of the proposed framework. 
The limitations of the proposed approach is that the graph representation model based on temporal random walk is not an inductive method, which can not obtain the representations for the newly added nodes. 
For future work, we plan to develop more specific graph representation methods to investigate various predictive tasks on Ethereum and extend the current framework to analyze other large-scale temporal or domain-dependent networks.


\end{document}